\documentstyle[aps,preprint,epsf]{revtex} 
\tighten 

\newcommand{\beq}{\begin{equation}} 
\newcommand{\eeq}{\end{equation}} 
\newcommand{\bea}{\begin{eqnarray}} 
\newcommand{\eea}{\end{eqnarray}} 
\newcommand{\nn}{\nonumber} 
\newcommand{\junk}[1]{} 
\def\<{\langle} 
\def\>{\rangle} 
\def\d{\partial} 
\def\+{\dagger} 
\def\U1A{U(1)$_{A}$} 
\def\LambdaQCD{\Lambda_{\rm QCD}}

\def\IIbar{I \bar{I}} 
\def\la{\langle} 
\def\ra{\rangle} 
\def\sech{{\mathrm sech}}
\def\MeV{\mathrm{MeV}}
\def\GeV{\mathrm{GeV}}
 
\begin{document} 
 
\title{Classical stability of \U1A domain walls \\ in dense matter QCD} 
\author{Kirk~B.~W.~Buckley} 
\address{Department of Physics and Astronomy, \\
University of British Columbia,\\ 
Vancouver, BC V6T 1Z1 Canada \\ 
Email: kbuckley@physics.ubc.ca} 
\maketitle 
 
\begin{abstract} 
It was recently shown that there exists metastable \U1A 
domain wall configurations in high-density QCD ($\mu \gg 1~\GeV$). 
We will assess the stability of such nontrivial 
field configurations at intermediate densities ($\mu < 1~\GeV$). 
The existence of such configurations at intermediate densities 
could have interesting consequences for the physics of neutron 
stars where such densities are realized.  
\end{abstract} 
 
 
\section{Introduction}   

In general, there do not exist domain walls or other topological defects
within the Standard Model. This is due to the trivial topology of the 
vacuum manifold. In contrast, these objects are quite common in  
condensed matter physics and cosmology. 
However, it has been realized only recently
\cite{ssz,sonKaon,fzstrings,kaplan} that topological defects 
such as domain walls and strings may exist within the Standard 
Model at a large chemical potential. 

It is well known that topological defects result from symmetries being 
broken. In the past few years there has been renewed interest in high 
density QCD. Similar to the BCS pairing in conventional superconductivity, 
the ground state of QCD at high density is unstable due to the formation
of a diquark condensate \cite{baillov,colorsc,colorsc2} 
(see \cite{cscqcdrev} for a review). 
In this new ground state, various symmetries which are present 
at $\mu=0$ are broken by the presence of this non-zero diquark condensate.
This leads to the formation of the various topological defects discussed
in \cite{ssz,sonKaon,fzstrings,kaplan}.

In \cite{ssz}, it was shown that at high densities ($\mu \gg 1~\GeV$) 
there exist domain wall solutions which interpolate between the same vacuum  
state, in which the \U1A phase of the diquark condensate
varies between $0$ and $2\pi$. This type of domain wall which  
interpolates between the same vacuum state has been studied  
before in the context of axion models \cite{axiondw}. 
It is interesting to note that similar domain wall configurations are
present for the zero density case in the large $N_c$ limit and
could be present for the physically relevant case of $N_c=3$ \cite{fz}, 
in which case they can be studied at the BNL Relativistic Heavy
Ion Collider (RHIC) \cite{shurzhit}. 
Given this, one might ask 
what happens between these two regions of $\mu=0$ and 
$\mu \gg 1~\GeV$. The main goal of the present paper is the analysis
of the classical stability of the \U1A domain walls for $\mu \lesssim 1~\GeV$. 
We demonstrate that the \U1A domain walls are classically stable down to  
densities $\mu \simeq 800~\MeV$. The main idea is to interpolate between
$\mu \gg 1~\GeV$ (where perturbation theory is justified) and 
$\mu \simeq \mu_c$
(where instanton calculations lead to a reasonable description).
Ideally, we would like to be able to make 
definitive statements on the stability of such configurations all the way 
down from large $\mu$ to $\mu_c \sim 500 ~\MeV$, where $\mu_c$ is the critical 
chemical potential above which the color superconducting phase occurs.  
Unfortunately, 
we are unable to make a definitive statement for 
$\mu_c \lesssim \mu \lesssim 800~\MeV$
due to a lack of theoretical control in this region. 
One can only speculate on the behavior in this region. 

This work is organized as follows. In Sec. 2, we will construct the effective 
potential. In Sec. 3, we will describe the nontrivial domain wall 
solutions. In Sec. 4, we will examine the classical stability of such  
configurations under small perturbations. In Sec. 5, we will end with  
concluding remarks and future considerations.  
 
 
\section{The Effective Potential} 
 
It is well known that the ground state of $N_f=2,3$, $N_c=3$ QCD exhibits  
the Cooper pairing phenomenon as in conventional superconductivity (known 
as the color 2SC (2 flavor color superconducting) and CFL 
(color-flavor locked) phases of QCD \cite{colorsc,colorsc2,CFL,rapp}). 
In what follows, we will consider the $N_c=N_f=3$ CFL phase. 
In the CFL phase, the condensates take the form
\bea 
\label{qqCFL} 
  \<q^{ia}_{L\alpha} q^{jb}_{L\beta} \>^* &\sim&   
  \epsilon_{\alpha\beta\gamma} \epsilon^{ij}\epsilon^{abc} X_c^{\gamma} , 
  \nonumber \\ 
  \<q^{ia}_{R\alpha} q^{jb}_{R\beta} \>^* &\sim&   
  \epsilon_{\alpha\beta\gamma} \epsilon^{ij}\epsilon^{abc}Y_c^{\gamma} , 
\eea 
where $L$ and $R$ represent left and right handed quarks; $\alpha$, $\beta$, 
and $\gamma$ 
are the flavor indices; i and j are spinor indices; a, b, and c are  
color indices; and $X_c^{\gamma}$ and $Y_c^{\gamma}$ are the condensates
which are complex color-flavor matrices. 
As the magnitudes of these condensates depend on the color 
index $c$, one can easily see that these objects are not  
gauge invariant by themselves. In order to construct a gauge invariant field,  
the following matrix which describes the octet of mesons and axial singlet
was considered in \cite{ss,ss2}: 
\beq 
\label{compsigma}
\Sigma_{\gamma}^{\beta}=X Y^{\+} = \sum_c X_c^{\beta} Y^{c *}_{\gamma}. 
\eeq 
If a \U1A rotation ($q \rightarrow \exp^{i \gamma_5 \alpha/2} q$) 
of this gauge-invariant field $\Sigma$ is performed, 
we see that the fields (\ref{qqCFL}) transform as
\bea 
X \rightarrow e^{-i \alpha} X, \nonumber \\ 
Y \rightarrow e^{+i \alpha} Y, 
\eea 
and therefore 
\beq 
\label{sigma}
\Sigma \rightarrow e^{-2 i \alpha} \Sigma. 
\eeq 
Goldstone's theorem states that there must be a single Goldstone mode $\eta'$
associated with the breaking of this symmetry. Given this, we can 
parametrize the field as follows:
\beq
\label{sigmadef}
\Sigma=\Sigma_o~e^{\rho}~e^{-i \phi},
\eeq
where the phase $\phi= \eta'/f$ is defined as a dimensionless
field which describes the $\eta'$ boson, $f$ is the corresponding 
decay constant, $\Sigma_o$ is the vacuum expectation value of the 
composite field (\ref{compsigma}),
and $\rho$ is another dimensionless field which 
describes the fluctuations of the magnitude of the 
condensate\footnote{The field $\rho$ should not be confused with the familiar 
$\rho$ meson in QCD.} (analogous
to the $\sigma$ field related to the fluctuations of the 
$\la \bar{q} q \ra$ chiral condensate). We choose to parametrize the field 
as $e^{\rho}$ for convenience later on. 

In order to construct an effective potential describing the dynamics  
of the phase of the condensate $\phi$ as well as the magnitude $|\Sigma|$,  
there are two types of terms which must be included.  
The first term which explicitly breaks the \U1A symmetry
 was calculated in \cite{ssz,tytgat}  
by substituting the form of condensates given above into the  
instanton induced four-fermion Lagrangian \cite{thooft,svz}: 
\beq
\label{V1inst} 
V_{\mathrm{1-inst}}(\rho,\phi)=-a \mu^2 \Delta_o^2~e^{\rho} \cos \phi, 
\eeq 
where $\Delta_o$ is the value of the gap in the quark spectrum. 
The perturbative form of the expectation value of the
condensate has been used in arriving at this result \cite{ssz,son}:
\beq
\label{sigmao}
\la |\Sigma| \ra 
     \equiv \Sigma_o= \frac{9}{8 \pi^2} \frac{\mu^4 \Delta_o^2}{g^2}.
\eeq
For $N_f=3$, the dimensionless coefficient $a$ was found to be
\beq
\label{coeffa}
a= 7 \times 10^3 \left( \frac{m_s}{\mu} \right)
   \biggl(\ln{\mu\over\LambdaQCD}\biggr)^7
   \biggl({\LambdaQCD\over\mu}\biggr)^9,
\eeq
where $m_s$ is the mass of the strange quark. 
The mass of the corresponding $\eta'$ boson can be easily calculated by 
expanding the potential, $m_{\eta'}=2 \pi \sqrt{a}~\Delta_o$. According to 
\cite{ssz}, this potential is only under theoretical control when the 
mass of the $\eta'$ boson is much less than the typical scale for 
higher excitations $2 \Delta$, which corresponds to $a \ll 1/\pi^2$.
For physical values of the strange quark, this corresponds to a chemical 
potential of about $\mu \approx 700~\MeV$ (for $a \sim 1/\pi^2$). 

To be able to draw any conclusions about the stability of the domain wall 
solution for intermediate densities $(\mu_c \lesssim \mu \lesssim1~\GeV)$, 
one must include  
degrees of freedom which are related to the fluctuations of the 
absolute value of $|\Sigma|$. We do not know the effective potential 
in the region of interest; however, for qualitative discussions we shall
use a potential derived for asymptotically large $\mu$.   
The second type of term which must be included in an effective potential  
description is one which uniquely fixes the magnitude of the condensate.  
An effective potential for the
magnitude of the condensate $|\Sigma|$ was derived in \cite{miransky}  
in the perturbative region where the analytical form of the gap is known.  
This effective potential fixes uniquely the value of the vacuum condensate.  
The potential is of the Coleman-Weinberg type \cite{cw} and is 
given as follows: 
\bea 
\label{Vpert} 
V_{\mathrm{pert}}(|\Sigma|)&=& - \left( \frac{2 \nu \pi}{ \mu} \right)^2  
|\Sigma| 
      \left(1- \ln \frac{|\Sigma|}{\Sigma_o}   \right), \nn \\ 
  &=& -  \frac{\mu^2 \Delta_o^2}{ \pi^2} 
      e^{\rho}  (1 - \rho ), \\  
\nu&=& \sqrt{\frac{8 \alpha_s}{9 \pi}}, \nn
\eea 
where $\alpha_s$ is the standard strong coupling constant and the 
perturbative result of the condensate (\ref{sigmao}) has been used. 
In all calculations that follow we will assume $\Delta_o=100~\MeV$ 
as the numerical value for the gap. It should  
be noted that the potential (\ref{Vpert}) is justified only in the  
region 
\beq 
\label{pertjust} 
 \nu \log (|\Sigma| / \Sigma_o) \ll 1.  
\eeq 
Since we are considering $|\Sigma|\approx \Sigma_o$, the use of this 
effective potential is justified. 

We are interested in the region $\mu_c \lesssim \mu \lesssim 1~\GeV$, 
where Eq.~(\ref{Vpert}) is not literally correct. 
Even though the results stated above 
are not necessarily under theoretical control for $\mu \gtrsim \mu_c$, 
one can speculate on how the coefficients behave at intermediate densities
when $\mu_c \lesssim \mu \lesssim 1~\GeV$. Due to the fact that all  
the same symmetries are present as the chemical potential is lowered until 
reaching $\mu_c$, we expect the qualitative form of the effective  
potential (\ref{V1inst}) and (\ref{Vpert}) to remain the same. 
As the chemical potential is lowered, eventually perturbative calculations
which are valid at asymptotically large $\mu$ are no longer the correct 
description and instanton calculations become relevant. One would 
expect that these calculations must match up at some point. 
However, the coefficients in front  
of the potential could possibly be very different in the density  
region of interest. We will refer to the coefficient 
in front of the one-instanton potential as $\beta_1$ and the 
coefficient in front of the perturbative potential as $\beta_2$. 
The value of $\beta_1$ is essentially fixed by the form of the condensate
and by the constituent quark mass. Below the critical chemical 
potential $\mu_c$ at which the chiral phase transition occurs, this 
coefficient is independent of $\mu$ \cite{rapp}. Therefore, we would 
expect that $\beta_1$ would be some smooth function of $\mu$ which 
reaches its maximum value at $\mu= \mu_c$. 
At $\mu_c \lesssim \mu \lesssim1~\GeV$, the coefficient 
$\beta_2$ is modified by the formation of instanton-antiinstanton ($\IIbar$) 
molecules \cite{rapp}. We should also note that the coefficient 
$\beta_1$ can also be estimated from the instanton liquid model 
\cite{rapp} using the average size of instantons as well as 
requiring a constituent quark mass of about $350-400~\MeV$. 

\begin{figure}
\begin{center}
\vspace{1cm}
\epsfysize=3.0in
\epsfbox[ 33 523 326 725]{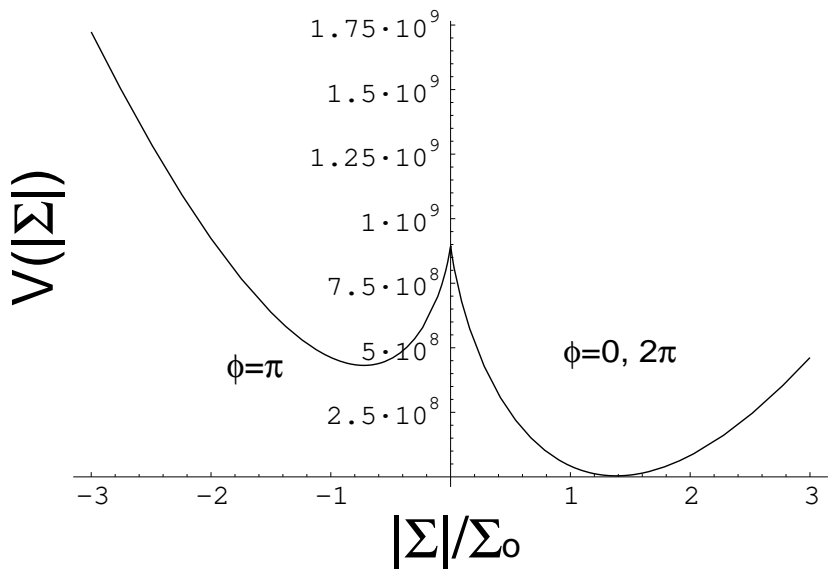}
\end{center}
\caption{ The cross section of the effective potential is shown 
above for $\mu=800~\MeV$. On the left half 
$\phi=\pi$ is shown and on the right half $\phi=0,2\pi$ is shown.}
\label{potentialfig}
\end{figure}

Combining both terms we have an effective potential which is given by 
the following:
\beq
\label{Veff}
V(\rho,\phi)= - \beta_2 e^{\rho}  (1- \rho )
     - \beta_1 e^{\rho} \cos \phi.
\eeq
The values of the coefficients $\beta_1$ and $\beta_2$ are known for 
asymptotically large $\mu$:
\bea
\label{beta12}
\beta_1&=& a \mu^2 \Delta_o^2,         \\
\beta_2&=& \frac{\mu^2 \Delta_o^2}{\pi^2}. 
\eea
In Fig.~\ref{potentialfig}, we show the cross section of the potential 
at $\phi=0,2\pi$ and $\phi= \pi$ for  $\mu=800~\MeV$. 
Notice that the 
existence of an absolute minima at $\phi=0,2\pi$ and a saddle point
at $\phi= \pi$ allows for nontrivial configurations which wind around the 
barrier at $|\Sigma|=0$. In the limit $\mu \rightarrow \infty$, the 
parameter $a \rightarrow 0$ and the potential has degenerate minima
at $|\Sigma|=\Sigma_o$. The kinetic term is given by
\beq
\label{ke}
\frac{|\d_{o} \Sigma|^2 - u^2|\d_{i} \Sigma|^2}{|\Sigma|^2} =
  (\d_{o} \rho)^2 -u^2(\d_{i} \rho)^2 +(\d_{o} \phi)^2 -u^2(\d_{i} \phi)^2,
\eeq
where $u$ is the velocity which is different from $1$. The perturbative
values for the decay constant $f$ and velocity $u$ were calculated 
in \cite{bbs}.  
In order to fix the correct dimensionality, we must multiply the 
kinetic term by the appropriate powers of the decay constant. The full
effective Lagrangian up to two derivatives in the fields is then given by
\beq
\label{Lfull}
{\cal L}= f^2[(\d_{o} \rho)^2 -u^2(\d_{i} \rho)^2]
     +f^2[(\d_{o} \phi)^2 -u^2(\d_{i} \phi)^2] - V(\rho,\phi).
\eeq
In the above we have assumed that $f_{\eta'} \simeq f_{\rho} = f$. The exact 
numerical value for $f_{\rho}$ is not known. However, in the large $\mu$ 
limit $f_{\eta'} \sim f_{\rho} \sim \mu$ in order to have an appropriate 
scale for $m_{\rho} \sim \Delta_o$ (once again, the $\rho$ field 
should not be confused with the well known $\rho$ meson in QCD at $\mu=0$).


\section{Domain wall solutions}

As was done in \cite{ssz}, if we replace the field 
$|\Sigma|= \Sigma_o e^{\rho}$ by its vacuum expectation value (which is 
justified for $\mu \gg 1~\GeV$), 
the resulting potential is of the sine-Gordon
type. The Lagrangian is given by the following:
\beq
\label{LSG}
{\cal L}=f^2 [(\d_0\phi)^2 - u^2 (\d_i\phi)^2] -
      V_{\mathrm{1-inst}}(\phi),
\eeq
where the constant term has been dropped. The static domain wall solution
to the corresponding equation of motion is well known. Considering a 
domain wall in the $z$ direction, the solution is given by
\beq
\label{SGdw}
\phi_o(z) = 4 \arctan \left( \exp(-mz/u) \right),
\eeq
where $m$ is the mass of the $\eta'$.
This solution interpolates between the same vacuum state; at 
$z= \pm \infty$ we have $\phi=0,2\pi$, respectively. It is well known that 
this solution is absolutely stable under small perturbations 
$\phi =  \phi_o + \delta \phi$. In other words, the Schr\"odinger-type 
equation obtained by varying the field and 
linearizing the equation of motion,
\beq
\label{delphistab}
- \d_z^2~\delta \phi_n 
   + \frac{m^2}{u^2} \left(1 - 2~\sech^2\left(\frac{m}{u} z\right) \right)
  \delta \phi_n = \omega_n^2 \delta \phi_n,
\eeq
has the lowest eigenvalue $\omega_o=0$ corresponding to 
$\delta \phi_o = d \phi_o(z)/dz \sim \sech (mz/u) $. 
This is just the zero mode which is a result of translational invariance 
$z \rightarrow z + z_o$. Since the lowest eigenvalue is 
non-negative, the domain wall solution is stable under small
perturbations. It turns out that this is the only bound state 
which is a solution of Eq.~(\ref{delphistab}). 

In the case in which the replacement 
$|\Sigma| \rightarrow \Sigma_o$ is not done,
the solution must be modified. If we want to study a stable solution for $\mu$
which is not asymptotically large, we must include fluctuations in $\phi$ 
as well as in the $\rho$ direction (i.e., the absolute value of $\Sigma$). 
The two equations of motion for static solutions are given by
\bea
\label{eomrho}
2 f^2 u^2 \nabla_i^2 \rho &=& 
	\beta_2~\rho ~e^{\rho} 
          - \beta_1~e^{\rho}  \cos \phi, \\
\label{eomphi}
2 f^2 u^2 \nabla_i^2 \phi &=& 
          \beta_1~e^{\rho}  \sin \phi. 
\eea
Although we do not know the exact solution for this set of coupled
nonlinear differential equations, if 
$\beta_1/ \beta_2 < 1$ we can approximate the solutions.
In this case, the approximate solutions can be parametrized by
\bea
\label{approxsoln}
e^{\rho_o} &\approx& 1+\alpha \cos \phi_o, \\
\phi_o &\approx& 4 \arctan \left( \exp(-mz/u) \right),
\eea
where $\alpha \approx \beta_1/ \beta_2$. Our stability analysis will be 
based upon these approximate solutions of the equations of motion. 
Since the above solutions do not correspond to the exact solutions to the 
equations of motion (minimum energy path which winds around that barrier 
at $|\Sigma|=0$), there will be nonzero linear terms when the energy
of the system is perturbed about the domain wall solutions $\phi_o$ and
$\rho_o$. These will be estimated in the following section where the 
stability analysis is performed. 

 
\section{Stability analysis of domain walls} 
 
Although these domain walls may exist as classically stable objects at 
large densities, it is not immediately obvious if this is the case at 
intermediate densities. In order to examine the classical stability of the 
domain wall configurations in the region $\mu \lesssim 1~\GeV$,  
we must look at how the system reacts to small perturbations of the  
fields. The energy density  
is given by the following expression:
\beq
\label{Ephirho}
E(\phi,\rho) = \int_{-\infty}^{+\infty} dz [ f^2 u^2 (\nabla \rho)^2
 + f^2 u^2 (\nabla \phi)^2 + V(\phi,\rho) ].
\eeq
In order to express all integrals as dimensionless quantities, we 
will perform the following change of variables:
\beq
z'= \frac{z}{\lambda},~~\lambda=\frac{u}{m_{\eta'}}.
\eeq
The energy density is now given by
\beq
\label{Ephirhodim}
E(\phi,\rho) = \frac{f^2 u^2}{\lambda} \left[\int dz'  \left(
(\nabla \rho)^2  + (\nabla \phi)^2 
+ \frac{\lambda^2}{f^2 u^2} V(\phi,\rho) \right )\right],
\eeq
where the derivative is now taken with respect to the dimensionless coordinate
$z'$. We can see that this has the correct dimensions of $\MeV^3$ times 
the dimensionless integral in square brackets. 
Following the standard method for analyzing the stability of  
a classical solution which was briefly described in the previous 
section, we will expand the fields about their vacuum  
expectation values: 
\bea 
\label{fluc}
\rho \rightarrow \rho_o + \delta \rho,  \nn  \\ 
\phi \rightarrow \phi_o + \delta \phi. 
\eea
Next, we substitute Eq.~(\ref{fluc}) into Eq.~(\ref{Ephirho}) and perform 
an expansion, neglecting any terms greater that quadratic order in 
$\delta \rho$ and $\delta \phi$:
\bea
\label{Eexpansion}
E &\simeq& E^{(0)} + E^{(1)} + E^{(2)}  \nn \\
&\simeq& E(\phi_o,\rho_o) 
+ \gamma \int dz' ~\delta\rho \left( -2 \nabla^2 \rho_o 
	+ \left. \frac{\lambda^2}{f^2 u^2}
	\frac{\delta V}{\delta \rho} \right|_{\phi_o,\rho_o}\right)
\nn \\
&+& \gamma \int dz' ~\delta \phi \left( -2 \nabla^2 \phi_o 
	+ \left. \frac{\lambda^2}{f^2 u^2}
	\frac{\delta V}{\delta \phi} \right|_{\phi_o,\rho_o}\right) 
+   \gamma \int dz' ~\delta \rho \left( - \nabla^2  
	+ \left. \frac{\lambda^2}{2f^2u^2} \frac{\delta^2 V}{\delta \rho^2} 
	\right|_{\phi_o,\rho_o}\right) \delta \rho
\nn \\
&+&  
    \gamma \int dz' ~\delta \phi \left(\left. \frac{\lambda^2}{f^2u^2}
  	\frac{\delta^2 V}{\delta \phi \delta \rho} 
	\right|_{\phi_o,\rho_o} \right)\delta \rho
+   \gamma \int dz' ~\delta \phi \left( - \nabla^2  
	+ \left. \frac{\lambda^2}{2f^2u^2}\frac{\delta^2 V}{\delta \phi^2} 
	\right|_{\phi_o,\rho_o}\right) \delta \phi,
\eea
where $\gamma=f^2 u^2/ \lambda$. 
The first term $E^{(0)}$ in the above expansion is the energy density or wall
tension of the domain wall. 
In the case in which the domain wall solutions given by Eq.~(\ref{approxsoln})
were the exact solutions to the classical equations of motion, the linear
terms $E^{(1)}$ (proportional to $\delta \rho$ and $\delta \phi$) would be 
zero everywhere. Due to the fact that our solutions are not exact, these
must be considered. 

First, let us estimate the term which is linear in $\delta \phi$:
\beq 
\label{delphi1}
\delta \phi ^{(1)} = \frac{f^2 u^2}{\lambda}
	\int dz' \delta \phi \left( -2 \nabla^2 \phi_o 
	+ \frac{\lambda^2}{f^2 u^2} \beta_1 e^{\rho_o} \sin \phi_o \right).
\eeq
We know that $e^{\rho_o} \approx 1 + \rho_o$ and using the fact that 
$\phi_o$ is a solution to the equation of motion given by Eq.~(\ref{eomphi}) 
with $\rho=0$, we have:
\bea
\label{linearphi}
\delta \phi ^{(1)} 
  &\approx& \frac{f^2 u^2}{\lambda}
  \int dz' ~ \frac{\lambda^2}{f^2 u^2} \beta_1 \rho_o \sin \phi_o \delta \phi
  \nn \\
  &\approx& \frac{f^2 u^2}{\lambda}  
	\int dz' ~ a ~\pi^2  \sin 2\phi_o ~\delta \phi.
\eea
This linear term goes like $a \pi^2$ ($\sim 0.3$ for $\mu=800~\MeV$)
and the integrand is small compared to the wall tension $E^{(0)}$
 and can be neglected. 
The term which is linear in $\delta \rho$ is:
\beq
\delta \rho ^{(1)} = \frac{f^2 u^2}{\lambda} 
	\int dz' ~\delta \rho \left( -2 \nabla^2 \rho_o 
	+ \frac{\lambda^2}{f^2 u^2} \beta_2 \rho_o e^{\rho_o} 
	- \frac{\lambda^2}{f^2 u^2} \beta_1 e^{\rho_o} \cos \phi_o \right).
\eeq
Using again that $\rho_o \approx (\beta_1/ \beta_2) \cos \phi_o$ we see 
this simplifies to
 
\bea
\label{linearrho}
\delta \rho^{(1)} &\approx&  \frac{f^2 u^2}{\lambda}  
	\int dz' ~ \left(-2 a~ \pi^2
	\delta \rho  \nabla^2 (\cos \phi_o) \right).
\eea
Since this term is also proportional to $a \pi^2$, $\delta \rho \ll \rho_o$, 
and the integral of $ \nabla^2 (\cos \phi_o)$ is small, we 
can also neglect this term. The magnitude of the linear term
shows how far away we are from the exact solution. This information will 
be used in what follows for the stability analysis. 

Now we consider the most important quadratic term. A similar case 
involving two coupled scalar fields was looked at in \cite{2scalar}
and we will follow the standard procedure presented there closely. If the 
field configuration is classically stable, the second variation of the 
energy should be a positive differential operator. This means we must 
solve the following Schr\"odinger-type eigenvalue problem:
\beq
\label{schroeq}
   H\pmatrix{\delta \rho \cr \delta \phi} 
	= \omega^2 \pmatrix{\delta \rho \cr \delta \phi},
\eeq
where $H$ is the operator,
\beq
\label{Hop}
H= -\d_{z'}^2{\bf 1} + \frac{\lambda^2}{2 f^2 u^2} U,
\eeq
and {\bf 1} is the $2\times 2$ identity matrix. 
The potential $U$ is a $2\times 2$ matrix with elements:
\bea
\label{matrixpot}
U_{11}&=&\beta_2~(1+\rho_o) ~e^{\rho_o} 
          - \beta_1~e^{\rho_o}  \cos \phi_o , \\
U_{12}&=&U_{21}=\beta_1~e^{\rho_o}  \sin \phi_o , \\
U_{22}&=&\beta_1~e^{\rho_o}  \cos \phi_o .
\eea
If the domain wall solution is a stable one, then the operator $H$ is
positive-semidefinite. 
The eigenvalue equations can be decoupled
by diagonalizing the matrix $U$. We should note that only the potential 
term has to be diagonalized when looking for negative energy modes, as was
done in \cite{2scalar}. The result is
\bea
\label{eigenvalues}
U_{\pm}=\frac{1}{2} \left(a \pm \sqrt{a^2-4(ab-b^2-c^2)}  \right),
\eea
where $a$, $b$, and $c$ are defined as
\bea
\label{abc}
a&=&\beta_2~(1+\rho_o) ~e^{\rho_o}, \nn \\
b&=&\beta_1~e^{\rho_o}  \cos \phi_o, \\
c&=&\beta_1~e^{\rho_o}  \sin \phi_o. \nn
\eea
The operator $H$ now takes the following form:
\beq
\label{diag}
H = \pmatrix{-\d_{z'}^2 + \frac{\lambda^2}{2 f^2 u^2}U_+ & 0 \cr 0 & -
\d_{z'}^2 + \frac{\lambda^2}{2 f^2 u^2} U_-}.
\eeq
Since we can immediately see that $U_+ \geq 0$ for all $z'$ due 
to the fact that $a > 0$, there does not exist any negative eigenvalues
corresponding to the first equation in this transformed basis. Looking at 
the second eigenvalue equation, we have
\beq
\label{VminSE}
\left( -\d_{z'}^2 + \frac{\lambda^2}{2 f^2 u^2} U_- \right)\psi_n 
= \omega_n^2  \psi_n.
\eeq 
It is a well known theorem of quantum mechanics that there must 
exist at least one bound-state solution to Eq.~(\ref{VminSE}). 
Due to the fact that our domain wall solution should be invariant under 
translations in space $z' \rightarrow z' + z'_o$, there should be a 
corresponding zero mode in the spectrum of Eq.~(\ref{VminSE}). In the 
high density limit, we recover the familiar sine-Gordon equation and 
we know that there is only one bound state in the spectrum of 
Eq.~(\ref{delphistab}). If the exact solution to the equations of 
motion (\ref{eomrho}) and (\ref{eomphi}) were known, one would expect to 
see a corresponding mode with a vanishing eigenvalue in the spectrum of 
Eq.~(\ref{VminSE}). 
As the density is lowered ($\beta_1 \le \beta_2$) 
and the saddle point at $\phi = \pi$ is still present, 
one would expect the appearance of a mode with a negative eigenvalue 
corresponding to instability of the domain wall.  
Due to the fact
that there still must be a zero mode in the spectrum, 
the zero mode would become the first excited state of Eq.~(\ref{VminSE})
and the lowest mode would have some negative eigenvalue 
$\omega_0^2 < 0$ corresponding to the instability of the domain wall. The 
problem of stability analysis now reduces to determining the eigenvalues
corresponding to the bound states of Eq.~(\ref{VminSE}). The appearance 
of an additional bound state in the spectrum as the chemical potential 
is lowered will be the first sign that the system is approaching the 
point of instability. 

Since the solution corresponding to Eq.~(\ref{approxsoln}) is not the 
exact solution but does represent a path which winds around the 
barrier at $|\Sigma|=0$, it is quite possible that the zero mode 
could show up in the spectrum with a small nonzero eigenvalue. It would 
show up as a true zero mode only when the exact solution to the equation
of motion is substituted into Eqs.~(\ref{abc}). 

Although the potential $U_-$ is nontrivial, we will use a variational
approach in order to determine the upper bounds on $\omega_0^2$ and 
$\omega_1^2$. In choosing a trial wave 
function, we make the observation that the potential $U_-$ is quite 
similar to the same potential which arises when analyzing the stability of 
the sine-Gordon soliton, Eq.~(\ref{delphistab}). 
We will pick our normalized trial wave function accordingly:
\beq
\label{trial0}
\psi_0=\sqrt{\frac{\sigma}{2}}~\sech (\sigma z'),
\eeq
with $\sigma$ being the variational parameter. Note that this trial wave 
function satisfies the required boundary conditions 
$\psi_o(z'= \pm \infty) \rightarrow 0$. For the 
first excited state, we must pick an odd function of $z$. We will 
choose the properly normalized function
\beq
\label{trial1}
\psi_1=\sqrt{\frac{3 \sigma}{2}} \tanh(\sigma z')~\sech (\sigma z').
\eeq
Applying the variational principle, we must calculate
\beq
\label{evar}
E_n^{(2)}(\sigma)= \frac{f^2 u^2}{\lambda}
 \langle \psi_n | \left(-\d_{z'}^2 + \frac{\lambda^2}{2 f^2 u^2} U_- \right) 
    | \psi_n \rangle,
\eeq
and minimize this quantity with respect to $\sigma$ to obtain an upper bound
$E_n^{(2)}(\sigma)$ on the energy of the $n{\mathrm th}$ state. The integral 
given by Eq.~(\ref{evar}) must be performed numerically. 

\begin{figure}
\begin{center}
\vspace{1cm}
\epsfysize=3.0in
\epsfbox[ 25 536 332 734]{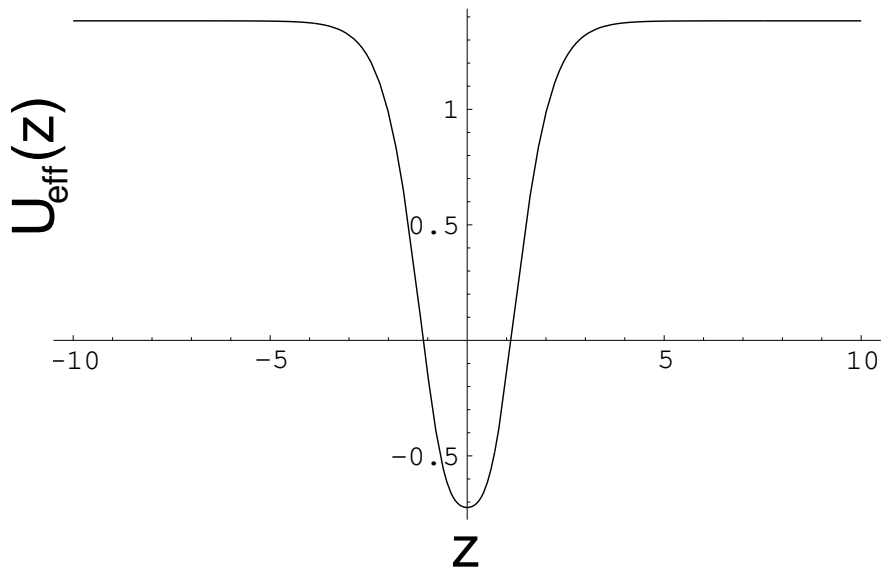}
\end{center}
\caption{ This figure shows the effective Schr\"odinger potential 
$U_{\mathrm{eff}}(z')=(\lambda^2/2f^2 u^2) U_-(z')$ as a function 
of the dimensionless coordinate $z'$ where $U_-$ is 
given by Eq.~(\ref{eigenvalues}). The presence of a negative eigenvalue
in the spectrum of this potential is indicative of the instability of the 
classical domain wall solution. The effective potential is shown in this 
figure for $\mu=800~\MeV$.}
\label{vminusfig}
\end{figure}

We now have the framework in place in 
order to test the stability of our domain wall configurations. We will assume
the perturbative value for the constants $f$ and $u$ as calculated in 
\cite{bbs}, 
$f^2=\mu^2/(8 \pi^2)$ and $u^2=\frac{1}{3}$. Setting 
$\mu=800~\MeV$ and $m_s=100~\MeV$, we see that the ratio of the coefficients 
is $\beta_1/ \beta_2 \approx 0.3$. In this case, the 
linear terms (\ref{linearphi}) and (\ref{linearrho}) 
are small and the solutions given 
by Eq.~(\ref{approxsoln}) are valid approximations to the exact solutions. 
In Fig.~\ref{vminusfig}, we show the effective Schr\"odinger potential 
$U_{\mathrm{eff}}=\lambda^2U_-/(2f^2u^2)$ for $\mu=800~\MeV$.
For the above choice of 
parameters, the wall tension given by Eq.~(\ref{Ephirhodim}) is
\beq
\label{walltension800}
E^{(0)}(\phi_o,\rho_o) \approx 17.48 \frac{f^2 u^2}{\lambda}. 
\eeq
For the trial wave functions given in Eqs.~(\ref{trial0}) and (\ref{trial1}), 
the following results for 
the two bound states of Eq.~(\ref{VminSE}) were obtained:
\bea
\label{results}
E_0^{(2)}(\sigma_{\mathrm{min}}) \le -0.016 \frac{f^2 u^2}{\lambda},\nn \\
E_1^{(2)}(\sigma_{\mathrm{min}}) \le +1.163 \frac{f^2 u^2}{\lambda}.    
\eea
From this, we can see that $E_1^{(2)} \gg E_0^{(2)}$ and both of these 
quantities are much less the wall tension given by 
Eq.~(\ref{walltension800}). Even though the ground state
energy seems to be negative, due to the fact that $E_1^{(2)} \gg E_0^{(2)}$ we 
can associate this mode with the zero mode. The small nonzero eigenvalue
is actually an artifact of our approximations. 
The appearance of a negative mode is merely a consequence of the approximate 
solutions (\ref{approxsoln}) as discussed above. 
This identification can be verified
by increasing the chemical potential. When the above calculations are 
repeated as the chemical potential is 
increased, we see that the energy $E_0^{(2)} \rightarrow 0$. This result is
expected due to the fact that the $|\Sigma|$ field can be integrated out
as $\mu$ increases and the sine-Gordon type theory is recovered. 
As $\mu$ is increased, we also see that the eigenvalue of the first excited
state $\omega_1$ increases towards the maximum value of $U_{\mathrm{eff}}$. 
Eventually, as $\mu$ is increased further, there is no longer a first excited
bound state in the spectrum. 
The calculation of $E_0^{(2)}$ and $E_1^{(2)}$ was done with the 
variational functions chosen to be different from 
Eqs. (\ref{trial0}) and (\ref{trial1}). The results obtained were the 
same order of magnitude as stated above (\ref{results}). This supports
our interpretation of the $E_0^{(2)}$ state as the would-be translational mode
if exact solutions are known. The magnitudes of the linear terms 
(\ref{linearphi}) and (\ref{linearrho}) are approximately the same order of 
magnitude of $E_0^{(2)}$, which supports our interpretation of $E_0^{(2)}$
as the zero mode.

 
\section{Conclusion} 

The main goal of this paper was an analysis of the classical stability 
of \U1A domain walls \cite{ssz}. Naively one would expect that decreasing 
$\mu$ from $\mu \gg 1~\GeV$ (when the calculations are under 
control \cite{ssz}), we inevitably face the situation in which the domain walls
become unstable objects due to the fast growth of the coefficient $\beta_1$ 
(\ref{Veff}). This naive expectation may not necessarily be correct due to
the even faster growth of the coefficient $\beta_2$ (\ref{Veff}), which 
receives contributions from the formation of $\IIbar$ molecules as well
as perturbative contributions. 

What we have actually demonstrated is that the domain wall solution remains
classically stable down to $\mu \simeq 800~\MeV$. In order to assess the 
stability of the domain walls for $\mu_c \lesssim \mu \lesssim 800~\MeV$,
one must explicitly include the $\IIbar$ contribution in the effective 
potential (which is expected to be the dominant contribution at 
$\mu \sim \mu_c$ \cite{rapp}). This would hopefully lead to
$\beta_1/\beta_2 < 1$, ensuring the classical stability of the \U1A 
domain walls. Unfortunately, due to the lack of information
in this region we cannot generalize our results to below $800~\MeV$ to $\mu_c$.
We can argue that the \U1A domain walls remain classically stable down
to $\mu \gtrsim \mu_c$ due to the faster growth of the coefficient $\beta_2$
compared to $\beta_1$ as $\mu$ is decreased. The ratio 
$\beta_1/\beta_2$ may be very sensitive to changes in the 
instanton size distribution or the various form factors. This is a difficult 
problem and unfortunately I do not know how to estimate such contributions
for a small chemical potential.

We should remark here that the stability of the \U1A domain wall implies a 
classical stability of the \U1A strings \cite{fzstrings}, 
which become the edge of the domain walls. 
It remains to be seen whether these (or other) topological defects 
will have any impact on the physics of neutron stars and other 
compact stellar objects with high core density $\mu$. 

\acknowledgements 
 
I would like to thank A. Zhitnitsky for suggesting this problem to me and  
for many helpful discussions and suggestions. I would also like to 
thank T. Sch\"afer for interesting discussions during the 
National Nuclear Physics Summer School in Bar Harbor, ME. 
This work was supported in part by the Natural Sciences 
and Engineering Research Council of Canada.

\end{document}